\documentclass[aps, prd, superscriptaddress, twocolumn,a4paper]{revtex4-1}
\usepackage{hyperref}
\usepackage{graphicx}
\usepackage[sort&compress]{natbib}
\usepackage{latexsym}
\usepackage{epsfig}
\usepackage[english]{babel}
\usepackage{graphicx}
\usepackage[T1]{fontenc}
\usepackage[utf8]{inputenc}
\usepackage{amsmath}

\usepackage{comment}
\usepackage{natbib}
\usepackage{bm}
\usepackage{amssymb}

\usepackage{mathtools}

\date{\today}

\begin{document}
\title{Fluctuations and thermodynamic geometry of the chiral phase transition}

\author{Paolo Castorina}
\affiliation{School  of  Nuclear  Science  and  Technology,  Lanzhou  University,
222  South  Tianshui  Road,  Lanzhou  730000,  China.}
\affiliation{Istituto Nazionale di Fisica Nucleare, Sezione di Catania, I-95123 Catania, Italy.}
\affiliation{Institute of Particle and Nuclear Physics, Faculty of Mathematics and Physics, Charles University, V Hole\v{s}ovi\v{c}k\'ach 2, 18000 Prague 8, Czech Republic.}

\author{Daniele Lanteri}\email{daniele.lanteri@ct.infn.it}
\affiliation{Istituto Nazionale di Fisica Nucleare, Sezione di Catania, I-95123 Catania, Italy.}
\affiliation{Dipartimento di Fisica e Astronomia, Universit\`a  di Catania, I-95123 Catania, Italy.}

\author{Marco Ruggieri}\email{ruggieri@lzu.edu.cn}
\affiliation{School  of  Nuclear  Science  and  Technology,  Lanzhou  University,
222  South  Tianshui  Road,  Lanzhou  730000,  China.}

\begin{abstract}
We study the thermodynamic curvature, $R$, 
around the chiral phase transition at finite temperature and chemical potential,
within the quark-meson model augmented with meson fluctuations.
We study the effect of the fluctuations, pions and $\sigma-$meson, 
on the top of the mean field
thermodynamics and how these affect $R$ around the crossover.
We find that for small chemical potential the fluctuations enhance the magnitude of $R$,
while they do not affect substantially the thermodynamic geometry in proximity of the critical endpoint.
Moreover, in agreement with previous studies
we find that $R$ changes sign in the pseudocritical region, suggesting a change of the nature of interactions at the mesoscopic
level from statistically repulsive to attractive.
Finally,
we find that in the critical region around the critical endpoint $|R|$ scales with the correlation volume,
$|R|=K\xi^3$ with $K=O(1)$, as expected from hyperscaling; 
far from the critical endpoint the correspondence between $|R|$
and the correlation volume is not as good as the one we have found at large $\mu$, which is not surprising
because at small $\mu$ the chiral crossover is quite smooth; nevertheless, we have found that $R$ develops a characteristic
groove structure, suggesting that it is still capable to capture  the pseudocritical behavior of the condensate.

\end{abstract}
\pacs{.s}
\maketitle

\section*{Introduction}
The thermodynamic theory of fluctuations allows to define
a manifold spanned by intensive thermodynamic variables, $\{\beta^k\}$ with $k=1,2,\dots,N$, and equip this with 
the notion of a distance, $d\ell^2=g_{ij}(\beta^1,\beta^2,\dots,\beta^N)d\beta^i d\beta^j$ where $g_{ij}$ is the metric tensor,
that depends in general of the $\{\beta^k\}$ and measures the probability of a fluctuation between two equilibrium states.
The metric tensor can be computed from the derivatives of the thermodynamic potential,
therefore the knowledge of the latter is enough to define the metric on the manifold.
Thermodynamic stability requires $g>0$ where $g$ is the determinant of the metric;
the condition $g=0$ determines a phase boundary in the $\{\beta^k\}$ space and $g<0$ corresponds
to regions of thermodynamic instability.

By means of $g_{ij}$ it is possible to define the scalar curvature, $R$, 
using the standard definitions of the Riemann geometry;
in this context, $R$  is named the thermodynamic curvature,
and the theory that studies $R$ is called thermodynamic 
geometry~\cite{Weinhold:1975get,Weinhold:1975gtii,Ruppeiner:1979trg,Ruppeiner:1981agt,Ruppeiner:1983ntp,Ruppeiner:1983tcf,
Ruppeiner:1985cgt,Ruppeiner:1985tcv,Ruppeiner:1986tcv,Ruppeiner:1990tig,Ruppeiner:1990tof,Ruppeiner:1991rgc,
Ruppeiner:1993afg,Ruppeiner:1995rgf,Ruppeiner:1998rgc,Ruppeiner:2005rgr,Ruppeiner:2008tcb,Ruppeiner:2010tci,Ruppeiner:2012tct,
Ruppeiner:2012tpw,Ruppeiner:2013tfb,Ruppeiner:2013tar,Ruppeiner:2014utg,Ruppeiner:2015trf,Ruppeiner:2015tsf,Ruppeiner:2015tsw,
Ruppeiner:2016smt,Ruppeiner:2017svc,Wei:2013ctb,Janyszek:1989rtm,Janyszek:1,Janyszek:2,
Sahay:2017gcb,Castorina:2018ayy,Mirza:2009nta,Mirza:2008rag,Castorina:2018gsx,Castorina:2019jzw,
Ruppeiner:maybe,covariant:evolution,geometrical:aspects,crooks:measuring,bellucci:PA,Zhang:2019neb}.
One of the merits of $R$ is that it carries the physical dimensions of a volume
and because of hyperscaling, around  a second order phase transition
$|R|\propto\xi^d$ where $d$ denotes the spatial dimension and
$\xi$ is the correlation length: as a consequence, $R$ diverges at a second order phase transition,
and by means of $R$ it is possible to estimate $\xi$
by virtue of pure thermodynamic functions. 
In general, the divergence of $R$ at a second order phase transition occurs in correspondence of 
the condition $g=0$, therefore looking for phase transitions in the $\{\beta^k\}$ space it is equivalent
to look for the zeros of $g$ or for the divergences of $R$; there are however other possibiliteis, like the divergence of one of the metric elements or of their derivatives (see eq.~\eqref{eq:bnn}).

In this study, we analyze the thermodynamic geometry,
and in particular the thermodynamic curvature, of the Quark-Meson (QM) model of Quantum Chromodynamics (QCD),
augmented with the fluctuations of the $\sigma-$meson and pions,
see~\cite{Ruggieri:2013cya,Ruggieri:2014bqa,Frasca:2011zn,Skokov:2010sf,Lenaghan:1999si,
Zacchi:2017ahv} and references therein.
Despite the abundant literature about thermodynamic curvature, a systematic study of the effect of fluctuations
on $R$ is missing, therefore we aim to fill this gap addressing the questions of how fluctuations influence the thermodynamic geometry.

Being this a first study on fluctuations in the context of thermodynamic curvature,
we introduce the fluctuations in the simplest way possible, namely using the Cornwall-Jackiw-Toumbulis (CJT) effective action
formalism for composite operators \cite{Cornwall:1974vz} and limiting ourselves to the largely used Hartree approximation
\cite{Lenaghan:1999si,Zacchi:2017ahv} in which momentum dependent self-energy diagrams are neglected.
Within these approximations, the effect of the interaction of the fluctuations with the medium is a shift in their mass
that can be computed solving self-consistently the Schwinger-Dyson equations for the propagators and for the
mean field condensate. 
Moreover, it is possible to write down in a simple form the    contributions of the fluctuations to the 
thermodynamic potential, therefore to evaluate the effect on the thermodynamic curvature.
Possible future improvements are mentioned briefly in the Conclusions.

Although this is a study about thermodynamic geometry, the model we use has been built up for modeling
the chiral phase transition of QCD at high temperature and finite chemical potential,
therefore it is useful to summarize briefly a few known facts about the QCD phase diagram 
and how this relates to that of the QM model, to give more context to the subject that we discuss here.
At zero baryon chemical potential,  from first principles Lattice QCD calculations we learn that
QCD matter experiences a smooth crossover from
a low temperature confined phase, in which chiral symmetry is spontaneously broken,
to a high temperature phase in which chiral symmetry is approximately
restored \cite{Borsanyi:2013bia,Bazavov:2011nk,Cheng:2009zi,Borsanyi:2010cj,Borsanyi:2010bp}.
Since the chiral restoration at large temperature is a smooth crossover, it is not possible to define
uniquely a critical temperature, rather it is more appropriate to define a pseudo-critical region,
namely a range of temperature in which several physical quantities (chiral condensate, pressure, 
chiral susceptibility and so on) experience substantial changes. 
This crossover is reproduced by the QM model, and the pseudo-critical temperature predicted by the model
is in the same ballpark of the pseudo-critical temperature od QCD, that is $T_c\approx 150$ MeV $\approx 10^{12}$ K.
At large finite baryon chemical potentials the
sign problem forbids reliable first principle calculations, therefore effective models like the QM model have been used
to study the phase structure of QCD at finite $\mu$ and it has been found that 
the smooth crossover becomes a first order phase transition if $\mu$ is large enough: this suggests the existence
of a critical endpoint (CEP) in the $(T,\mu)$ plane at which the crossover becomes a second order phase transition with divergent
susceptibilities, and this point marks the separation between the crossover on the one hand and the first order line
on the other hand. 
Recently, information theory has also been applied to the QCD phase diagram~\cite{Kashiwa:2020waa}.

We anticipate here the main results. The curvature is found to be positive at low temperature,
as for an ideal fermion gas; then a change of sign is observed near the chiral crossover, where $R$
develops a local minimum which becomes more pronounced when the chemical potential is increased; finally,
$R$ becomes positive again at high temperature and approaches zero from above.
A change of sign of $R$ has been observed for many substances
\cite{Ruppeiner:2008tcb,Ruppeiner:2012tct,Ruppeiner:2013tar,Ruppeiner:2013tfb,Ruppeiner:2015tsf,
Ruppeiner:2015tsw,Ruppeiner:2015trf,Ruppeiner:2016smt,Wei:2013ctb}
as well as in previous studied on the thermodynamic curvature of the chiral phase 
transition~\cite{Castorina:2019jzw,Zhang:2019neb}
and it has been interpreted in terms of the nature of the attractive/repulsive microscopic interaction.
We support this idea here, and we interpret the change of sign of $R$ around the chiral crossover as a rearrangement
of the interaction at a mesoscopic level, from statistically repulsive far from the crossover to attractive around the crossover.
Moreover, $|R|$ increases along the critical line as $\mu$ is increased from zero to the corresponding CEP value
and diverges at the CEP: this is in agreement with $|R|\propto\xi^3$ since the correlation
length remains finite at the crossover but increases as the crossover becomes sharper and eventually diverges at the critical endpoint.
We check quantitatively the relation between $R$ and $\xi$ near the CEP by identifying $\xi=1/M_\sigma$,
where $M_\sigma$ is the pole mass of the $\sigma-$meson that carries the fluctuations of the $\sigma$ field.
Even more, we find that fluctuations enhance $|R|$ at the crossover at small $\mu$, and we interpret this
as the fact that the fluctuations make the chiral broken phase more unstable and favor chiral symmetry restoration
at finite temperature; near the CEP we do not find substantial effects of the fluctuations on $R$,
and we interpret this as the fact that even without fluctuations, the mean field thermodynamic potential
predicts a second order phase transition at the CEP with divergent susceptibilities and a divergent 
curvature \cite{Castorina:2019jzw,Zhang:2019neb}, and the fluctuations cannot change this picture but
can only alter the values of the critical exponents.

The plan of the article is as follows. In Section~\ref{sec:TG} we briefly review the thermodynamic geometry
and in particular the thermodynamic curvature. In Section~\ref{sec:QM} we review the QM model.
 In Section~\ref{sec:Results} we discuss
$R$ for the QM model. Finally, in Section~\ref{sec:SC} we draw our conclusions.
We use the natural units system $\hbar=c=k_B=1$ throughout this article.

\section{\label{sec:TG}Thermodynamic geometry} 
Consider a thermodynamic system in the grand-canonical ensemble 
whose equilibrium state is characterized by the pair $(T,\mu)$, 
where $T$ is the temperature and $\mu$ is the chemical potential conjugated to particle density. 
In order to define the thermodynamic geometry it is convenient
to shift to new coordinates $X=(X^1,X^2)=(\beta,\gamma)$ with $\beta=1/T$ and $\gamma=-\mu/T$. 

It is well known that a thermodynamic system at equilibrium can fluctuate to another equilibrium state
characterized by different values of $X$, and the probability of this fluctuation can be computed
within the standard thermodynamic fluctuation theory.
In order to formulate this as well as the geometry of thermodynamics,
it is possible to define a metric space on the 2-dimensional manifold spanned by $(\beta,\gamma)$
introducing a distance between two points, analogously to what is done in Riemannian geometry. 
In particular, we define the distance $d\ell^2$ as~\cite{Ruppeiner:1979trg,Ruppeiner:1995rgf}
\begin{equation}
d\ell^2=g_{\beta\beta}d\beta d\beta+2g_{\beta\gamma}d\beta d\gamma+g_{\gamma\gamma}d\gamma d\gamma,
\label{eq:distance}
\end{equation}
where for a system with grand-canonical partition function $\mathcal{Z}$ we have put
\begin{equation}
g_{ij}\equiv\frac{\partial^2{\log\mathcal{Z}}}{\partial X^{i} \partial X^{j}}=
\frac{\partial^2\phi}{\partial X^{i}\partial X^{j}}\equiv \phi_{,ij}
\label{Eq.definition}
\end{equation}
with $\phi\equiv\beta P$, the pressure $P=-\Omega$, with $\Omega$ representing the thermodynamic potential per unit volume;
moreover $\phi_{,ij}$ denotes the second derivative of $\phi$ with respect to $i$ and $j$.

The distance in Eq.~\eqref{eq:distance}
can be connected to the theory of thermodynamic fluctuations as follows.
The probability of the system fluctuating from 
$X=(\beta,\gamma)$ to $X+dX=(\beta+d\beta,\gamma+d\gamma)$ 
is given by
\begin{equation}
dp\propto \sqrt{g} \exp\left(-\frac{d\ell^2}{2}\right)d\beta d\gamma,
\label{eq:dP}
\end{equation}
where $g$ is the determinant of the metric tensor defined in Eq.(\ref{Eq.definition}).
Large probability of a fluctuation corresponds to small $d\ell^2$, while small probability to large $d\ell^2$.  
Therefore, a large thermodynamic distance between two equilibrium states, $X$ and $X+dX$,
means  a small probability to fluctuate from $X$ to $X + dX$;
vice versa, a small distance implies a large probability of fluctuate.  
In this sense, Eq.~\eqref{eq:distance} measures the distance in the $(\beta,\gamma)$ plane
between two thermodynamic states in equilibrium.

Thermodynamic stability requires that $g_{\beta\beta}>0$ and $g>0$, while $g=0$ corresponds to a phase boundary
and regions with $g<0$ are unstable. The stability conditions ensure that $d\ell^2$ is a positive definite quantity.
The second derivatives are related to the fluctuation moments:
\begin{align}
\phi_{,ij}&=\langle(F_{i}-\langle F_{i} \rangle)(F_{j}-\langle F_{j} \rangle)\rangle,
\end{align}
where $F_{i}$ denotes the physical quantities conjugated to $X^{i}$ and $\langle\cdots\rangle$ is the standard 
ensemble average. In our case we have:
\begin{align}
\phi_{,\beta\beta}&=\frac{1}{V}\;\langle(U-\langle U\rangle)^2 \rangle,\\
\phi_{,\beta\gamma}&=\frac{1}{V}\;\langle(U-\langle U\rangle)(N-\langle N \rangle) \rangle,\\
\phi_{,\gamma\gamma}&=\frac{1}{V}\;\langle(N-\langle N \rangle)^2 \rangle,
\end{align}
where $U, N$ denote the internal energy and the particle number respectively. 

Equipped with a metric tensor in the $(\beta,\gamma)$ manifold, it is possible to define the Riemann tensor,
\begin{equation}
R^i_{~klm}=\frac{\partial\Gamma^i_{km}}{\partial x^l}-\frac{\partial\Gamma^i_{kl}}{\partial x^m}
+\Gamma^i_{nl}\Gamma^n_{km}-\Gamma^i_{nm}\Gamma^n_{kl},
\label{eq:rmts}
\end{equation}
where the Christoffel symbols are given by
\begin{equation}
\Gamma^i_{kl}=\frac{1}{2}g^{im}
\left(
\frac{\partial g_{mk}}{\partial x^l}+\frac{\partial g_{il}}{\partial x^k}-\frac{\partial g_{kl}}{\partial x^m}
\right).
\end{equation} 
Standard contraction procedure allows to introduce the Ricci tensor, $R_{ij}=R^k_{~ikj}$ and the
scalar curvature, $R=R^i_{~i}$ that in this context is called the thermodynamic curvature.
For a two-dimensional manifold the expression for $R$ simplifies considerably, namely~\cite{Ruppeiner:1995rgf}
\begin{equation}
R=\frac{1}{2 g^{2}}\left|\begin{array}{ccc}
\phi_{,\beta \beta} & \phi_{,\beta \gamma} & \phi_{,\gamma \gamma} \\
\phi_{,\beta \beta \beta} & \phi_{,\beta \beta \gamma} & \phi_{,\beta \gamma \gamma} \\
\phi_{,\beta \beta \gamma} & \phi_{,\beta \gamma \gamma} & \phi_{,\gamma \gamma \gamma}
\end{array}\right|,\label{eq:bnn}
\end{equation}
where $|\ddots|$ denotes the determinant of the matrix.
Notice that the curvature diverges for $g=0$ namely on a phase boundary, unless the determinant in the numerator
of Eq.~\eqref{eq:bnn} vanishes on the same boundary.

It has been postulated that $|R|\propto \xi^3$ in proximity of a second order phase transition,
where $\xi$ is the correlation length of the fluctuations of the order parameter~\cite{Ruppeiner:1979trg}. 
This relation is natural in the hyperscaling hypothesis because $R$ brings the physical dimension of a volume;
being based on scaling, this relation should be valid only in proximity of a second order phase transition.
It is remarkable that many independent theoretical calculations based on different models confirm this hypothesis~\cite{Ruppeiner:1979trg,Janyszek:1990,Janyszek:2,Ruppeiner:1995rgf};
therefore, not only the study of $R$ in the $(\beta,\gamma)$ twofold brings information about the phase transitions,
but it allows for an estimate of the correlation volume based only on the thermodynamic potential rather than 
computing correlators: this is one of the merits of the thermodynamic geometry.

It has also been suggested that the sign of $R$ conveys details about the nature of the interaction,
attractive or repulsive, at a mesoscopic level in proximity of the phase transition.
Within our sign convention,
$R<0$ for an attractive interaction while $R>0$ corresponds to a repulsive one. These interactions not only include
real interactions~\cite{Ruppeiner:2011gm,Ruppeiner:2012tct,May:2012,May:2013,Ruppeiner:2015trf}, but also the statistical attraction/repulsion that ideal quantum gases feel in phase space~\cite{Oshima:1999,Ubriaco:2016,Mirza:2008rag,Ubriaco:2013,MehriDehnavi:2020},
so an ideal fermion gas has $R>0$ due to the statistical repulsion of the Fermi-Dirac statistics while
an ideal boson gas has $R<0$ due to the statistical attraction of bosons.
The thermodynamic curvature is known to be identically zero only for the ideal classical gas.
Other fields of application concern Lennard-Jones fluids~\cite{May:2012,May:2013}, ferromagnetic systems~\cite{Dey:2011cs}, gravitational systems and Black Holes~\cite{Chaturvedi:2014vpa,Sahay:2017hlq,Sahay:2010tx,Ruppeiner:2007hr,Aman:2003ug,Shen:2005nu,Aman:2005xk,Ruppeiner:2008kd,Sarkar:2008ji,Bellucci:2011gz,Wei:2012ui,Wei:2015iwa,Sahay:2016kex,Ruppeiner:2018pgn,Yerra:2020oph,Yerra:2020tzg,Lanteri:2020trb}, strong interacting matter~\cite{Castorina:2018ayy,Castorina:2018gsx,Castorina:2019jzw,Zhang:2019neb} and others~\cite{Dey:2011cs,Chaturvedi:2014vpa,Sahay:2017hlq,Diosi:1988vx,Diosi:1985wy}.

\section{\label{sec:QM}The Quark-Meson model}
In this section, we review the quark-meson (QM) model in which fermions (in our context, quarks) 
interact with mesons (that are the $\sigma$-meson and the pions in our work), and is based on the Lagrangian density
\begin{equation}
\mathcal L = \mathcal L_m + \mathcal L_f
\;,
\end{equation}
with the mesonic and fermionic parts respectively given by
\begin{equation}
\begin{split}
\mathcal L_m 
=&
\text{Tr}\left[\left(\partial_\nu \Phi\right)^{\mathcal y} \left(\partial^\nu \Phi\right)\right] 
-
m^2\;\text{Tr}\left(\Phi^{\mathcal y}\Phi\right)  
-\\
&
-
\lambda\;\left[\text{Tr}\left(\Phi^{\mathcal y}\Phi\right) \right]^2 
+
h\;\sigma 
\end{split}
\end{equation}
and
\begin{equation}
\mathcal L_f =
\overline \Psi i\,\gamma^\mu \partial_\mu \Psi
-
2\,g\,\overline \Psi\Phi\Psi
\;.
\end{equation}
Here $\Phi$ is the matrix field
\begin{equation}
\Phi \equiv \frac{1}{2}\;\sigma\;\tau^0
+
\frac{i}{2}\,\overrightarrow \pi \cdot \overrightarrow \tau  
\;,
\end{equation}
with
$\tau^0$ the unity matrix and $\overrightarrow \tau=\left(\tau_1,\tau_2,\tau_3\right)$ the Pauli matrix,
$\overrightarrow \pi = \left(\pi_1,\pi_2,\pi_3\right)$ is an isotriplet of pion fields, $\sigma$ is the isosinglet field and $\Psi$ is  a massless isodoublet quark field. 

A common approximation, done in particular in the context of effective field theories
for the quark chiral condensate of QCD, is that of mean field in which the meson fields are replaced
by their uniform, time independent saddle point values $\sigma=f_\pi$ and $\overrightarrow{\pi}=0$.
In this study we want to go beyond the mean field approximation, including the quantum fluctuations of the meson fields
and studying their effect on the thermodynamic geometry
(the functional integral over the fermion fields can be done exactly on top of the mean field solution).
Within a gaussian approximation, the partition function of the model is given by
\begin{equation}
\mathcal Z = \mathcal Z_f\;\mathcal Z_m
\;,
\end{equation}
where the subscripts $f$ and $m$ stand for fermions and mesons respectively;
in this model, both quarks and meson fluctuations propagate on the background of 
the condensate of the $\sigma$ field, the value of which is determined consistently by solving
the gap equations (see below). 
The thermodynamic grand potential is 
\begin{equation}
\Omega = \Omega_f + \Omega_m 
\;.
\end{equation}
Before giving the expression for $\Omega$, 
we emphasize that both $\Omega_f$ and $\Omega_m$ contain ultraviolet divergent contributions arising from momentum integration
of the single particle energies, that correspond to the usual zero point energy of ideal gases of fermions and bosons;
these contributions cannot be simply subtracted since they contain a dependence from the condensate that in principle
affects the response of the condensate itself to temperature and chemical potential.  

We now give the expression of $\Omega$. 
Starting with $\Omega_f$, the standard renormalization procedure gives
\begin{equation}
\begin{split}
\Omega_f 
&=
\frac{g^4 N_c N_f}{8\;\pi^2} \sigma^4\ln \frac{Q_f}{g \sigma}
\\
&-
2 N_c N_f T\!\!\!
\int \!\!\!\!\frac{d^3 k}{(2\pi)^3}
\ln \left(1+e^{-\beta(\sqrt{k^2+g^2\,\sigma^2}-\mu)}\right)  
\\
&-
2 N_c N_f T\!\!\!
\int\!\!\!\! \frac{d^3 k}{(2\pi)^3}
\ln \left(1+e^{-\beta(\sqrt{k^2+g^2\,\sigma^2}+\mu)}\right)
\;.
\end{split}\label{eq:LLL}
\end{equation}
In the second and third lines of the right hand side of Eq.~\eqref{eq:LLL} 
we recognize the standard relativistic free gas thermodynamic potential at finite temperature and chemical potential;
on the first line of the right hand side of the same equation we show the zero temperature, zero chemical potential contribution
that is potentially divergent and has been renormalized
at the scale $Q_f$.

The mesonic contribution, $\Omega_m$, can be obtained via the standard
the CJT effective action formalism in the Hartree approximation in which momentum-dependent self-energy
corrections are neglected.  
Differently from \cite{Lenaghan:1999si,Zacchi:2017ahv} we do not include the vacuum term
of the meson potential, so the pressure of the pions and $\sigma-$meson is zero at $T=\mu=0$: 
the condensation energy takes contributions only from the classical potential plus the fermion loop,
while the mesons appear as excitation of the ground state at finite temperature.
This choice is done also for the sake of simplicity because 
including a further zero temperature, zero chemical potential renormalized term of the mesons would introduce
an additional renormalization scale that would lead to unexpected behaviors of the thermodynamic 
quantities \cite{Zacchi:2017ahv}. Within these approximations we have \cite{Lenaghan:1999si,Zacchi:2017ahv}
\begin{equation}\label{eq:Om}
\Omega_m 
=
\Omega_m^0
+
3\;B_{\pi}+B_{\sigma}
-
\frac{3\;\lambda}{4}
\;
\left(2\;A_{\pi}\;A_{\sigma}
+
5\;A_{\pi}^{2}
+
A_{\sigma}^{2}\right)
\;,
\end{equation}
 where
\begin{equation}
\Omega_m^0=\frac{m^2}{2} \;\sigma^2
+
\frac{\lambda\;\sigma^4}{4}
-h\;\sigma \;, \label{eq:OMEGAM0}
\end{equation}
is the mesonic part without fluctuations,
and for $\ell=\sigma,\pi$ we have put
\begin{equation}\label{eq:A}
\begin{split}
 A_\ell
 =& -
 \int\!\!\! \frac{d^3 k}{(2 \pi)^3} \! \frac{1}{E_\ell}\;\frac{1}{1-e^{\beta\;E_\ell}}
\;,   
\end{split}
\end{equation}
with  
$E_\ell = \sqrt{k^2 +M^2_\ell}$, and
 \begin{equation}
 \begin{split}
 B_\ell
 =
 2\; T\;\int \frac{d^3 k}{(2 \pi)^3} \;\ln\left(1-e^{-\beta\;E_\ell}\right)
 \;.
\end{split}
\end{equation}
Within this model, for given temperature and chemical potential the unknown are the value of the condensate, namely the expectation
value of $\sigma$, as well as the in-medium meson masses $M_\sigma$ and $M_\pi$:
these are obtained by solving the
gap equations, that are 
\begin{equation}\label{eq:GAP0}
\displaystyle
\begin{split}
h = &
\left[m^2
+
\lambda \,\sigma^2
+ 
3\;\lambda\;\left(A_\sigma+A_\pi\right)\right]\,\sigma
\\
&
-
\frac{g^4\;N_c\;N_f\;\sigma^3}{8\;\pi^2}
\;
\left(
1+4\;\ln \frac{g\;\sigma}{Q_f}
\right)
\\
&-
\frac{2\;N_c\;N_f}{\beta}\;\frac{\partial \Omega_{fT}}{\partial \Sigma}\Bigg|_{\overset{\scriptstyle\Sigma=\sigma}{\Pi=0}}, 
\;,
\end{split}
\end{equation}
\begin{equation}\label{eq:GAP0b}
\begin{split}
M_\sigma^2
=&
m^2
+
3\;\lambda\;\left(A_\pi+A_\sigma+\sigma^2\right)
\\
&-
\frac{g^4\;N_c\;N_f\;\sigma^2}{8\;\pi^2}
\;
\left(
7+12\;\ln \frac{g\;\sigma}{Q_f}
\right)
\\
&
-
\frac{2\;N_c\;N_f}{\beta}\;\frac{\partial^2 \Omega_{fT}}{\partial \Sigma^2}\Bigg|_{\overset{\scriptstyle\Sigma=\sigma}{\Pi=0}},
\;,
\end{split}
\end{equation}
\begin{equation}\label{eq:GAP0c}
\begin{split}
M_\pi^2
=&
m^2
+
\lambda\;\left(5\;A_\pi+A_\sigma+\sigma^2\right)
\\
&-
\frac{g^4\;N_c\;N_f\;\sigma^2}{8\;\pi^2}
\;
\left(1+4\;\ln \frac{g\;\sigma}{Q_f}
\right)
\\
&
-
\frac{2\;N_c\;N_f}{\beta}\;\frac{\partial^2 \Omega_{fT}}{\partial \Pi^2}\Bigg|_{\overset{\scriptstyle\Sigma=\sigma}{\Pi=0}}
\;,
\end{split}
\end{equation}
with
\begin{equation}
\begin{split}
\Omega_{fT}
=&
\int \frac{d^3 k}{(2\;\pi)^3}
\ln \left(1+e^{-\beta(\sqrt{k^2+g^2(\Sigma^2+\Pi^2)}-\mu)}\right)     
\\
&+
\int \frac{d^3 k}{(2\;\pi)^3}
\ln \left(1+e^{-\beta(\sqrt{k^2+g^2(\Sigma^2+\Pi^2)}+\mu)}\right)    
\;.
\end{split}
\end{equation}
The gap equations depend of the renormalization scale, $Q_f$, 
as well as of three parameters, $m$, $\lambda$ and $h$. 
At the tree-level, namely when no meson and quark loops are considered, 
the parameters $m$, $\lambda$ and $h$ are fixed to reproduce the physical values 
$m_\sigma$, $m_\pi$ as well as $\sigma =f_\pi$ at $T=0$ and $\mu=0$,
where we use small letters to denote physical masses at $T=\mu=0$;
without the fermion and meson loops these give
\begin{equation}
h\equiv h_{\mathrm{tree}}=m^2_\pi\,f_\pi\;, 
\end{equation}
\begin{equation}
m^2\equiv m^2_{\mathrm{tree}} = -\frac{m^2_\sigma-3\,m^2_\pi}{2} -\frac{f_\pi^2\;g^4\;N_c\;N_f}{4\;\pi^2} \;,  
\end{equation}
\begin{equation}
\lambda\equiv \lambda_{\mathrm{tree}} =\frac{m^2_\sigma-m^2_\pi}{2\;f^2_\pi} \;,
\end{equation}
where the subscript tree reminds that these are quantities computed using the tree-level potential.
In order to fix the renormalization scale we have to adopt one renormalization condition,
that is
\begin{equation}\label{eq:p0}
\lambda=\lambda_\mathrm{tree},
\end{equation}
where  $\lambda$ results from the
gap equations at $T=\mu=0$, namely
\begin{equation}\label{eq:p}
\lambda = 
\frac{m_\sigma^2-m_\pi^2 }{2\;f_\pi^2}
	+
	\frac{g^4\,N_c\,N_f}{8\,\pi^2}\,\left(3+4\,\ln \frac{g\,f_\pi}{Q_f}\right)
\;.
\end{equation}
$m^2$ and $h$ from the
gap equations at $T=\mu=0$ are always equal to the tree value:
\begin{equation}\label{eq:pb}
m^2
=
m^2_{tree}
\;,\qquad 
h 
=
h_{tree}\;.
\end{equation}
Finally, from eq.s~\eqref{eq:p0} and~\eqref{eq:p} we have 
\begin{equation}\label{eq:caso2a}
Q_f
=
e^{3/4}\;f_\pi\;g
\;.
\end{equation}

\section{\label{sec:Results}Results}
In this section we report and discuss our results. Firstly, we show briefly the effect of fluctuations
on the condensate, then we focus on the thermodynamic geometry. Our purpose is to show the existence
of a pseudo-critical region in which the condensate substantially decreases with temperature, then study the
elements of the thermodynamic metric as well as the scalar curvature around this region.
For the parameters  we take $f_\pi=93$ MeV, $m_\sigma=700$ MeV, $m_\pi=138$ MeV and finally
$g=3.6$: the latter is chosen so that the constituent quark mass at $T=\mu=0$ is $M=335$ MeV.
The resulting value of the renormalization scale is $Q_f	 = 709$ MeV.

\subsection{The condensate and the meson masses}
\begin{figure}[t!]
	\centering
	\includegraphics[width=0.8\columnwidth]{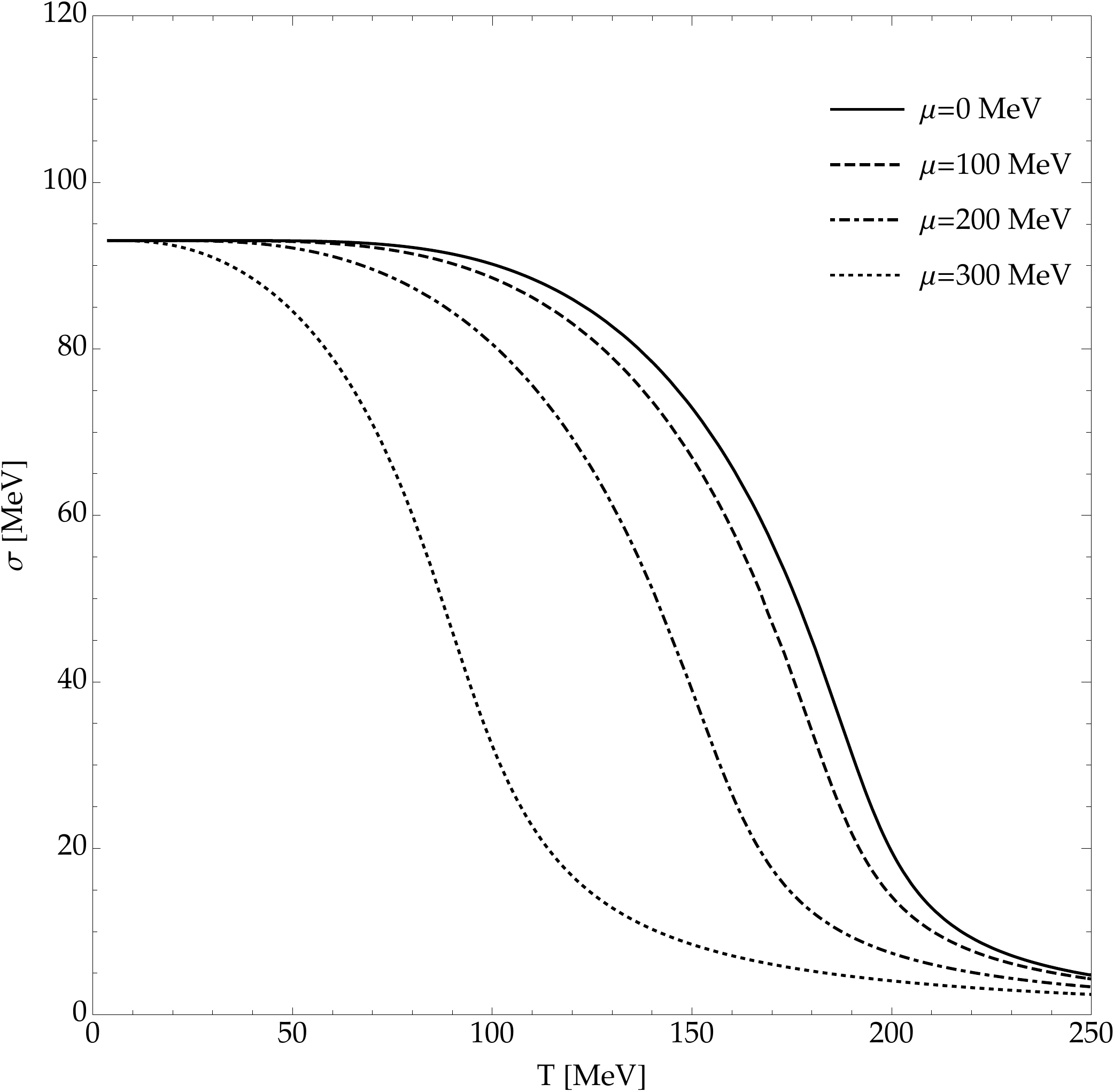}\\
	\includegraphics[width=0.8\columnwidth]{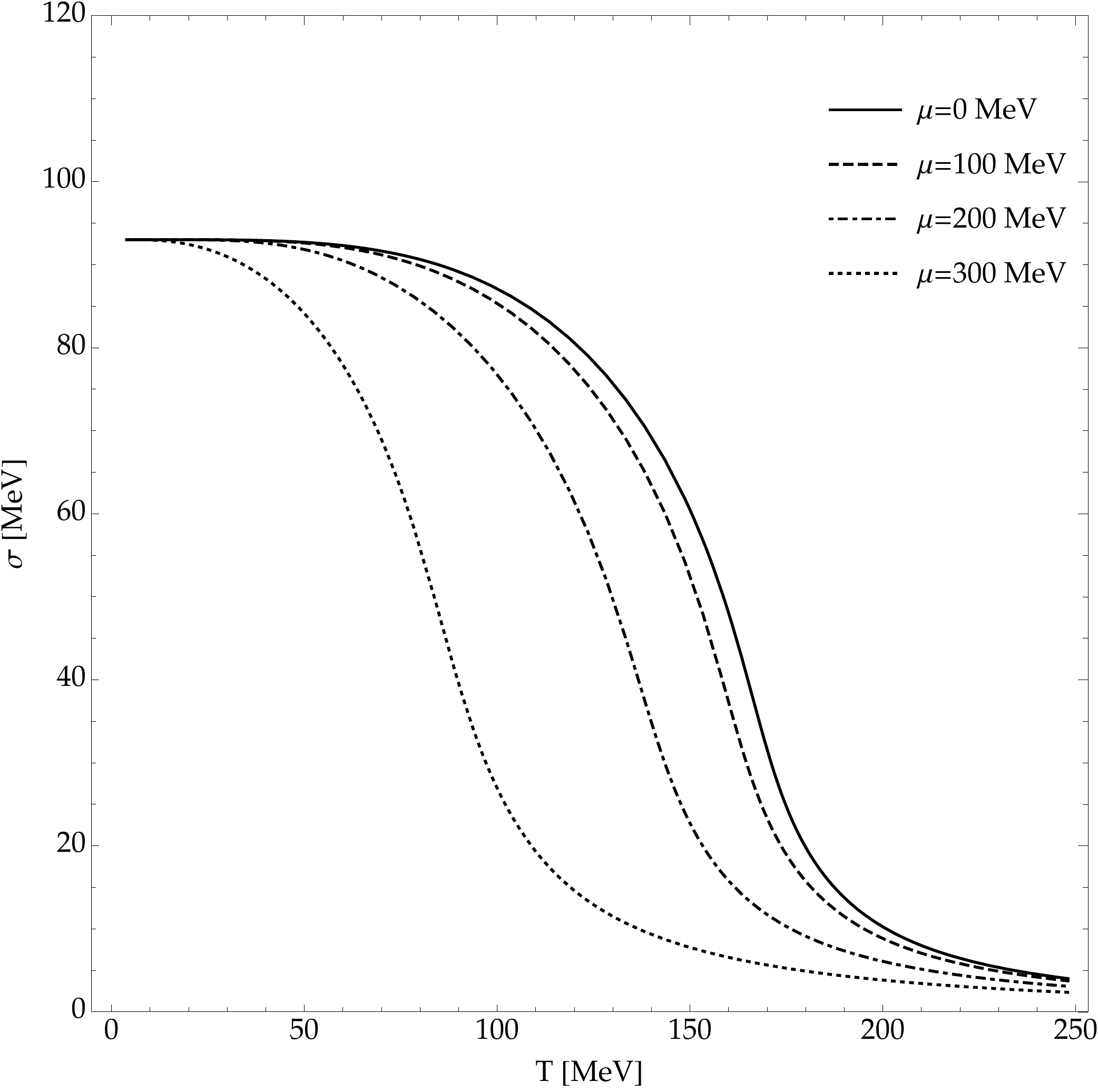}
	\caption{Condensate, $\sigma$, as a function of $T$ and for different values of the chemical potential.
	Upper panel corresponds to the case in which meson fluctuations are neglected, lower panel to the case in which
	meson fluctuations are included.}
	\label{fig:sigmaT}
\end{figure}

In Fig.~\ref{fig:sigmaT} we plot the condensate, $\sigma$, 
as a function of $T$ for several values of the chemical potential: $\mu=0$~MeV (continuous line), $\mu=100$~MeV (dashed), $\mu=200$~MeV (dot-dashed) and $300$~MeV (dotted).
The upper panel corresponds to the case in which meson fluctuations are neglected  (in this case the thermodynamic potential is $\Omega^0=\Omega_f+\Omega_m^0$, with $\Omega_f$ and $\Omega_m^0$ in eq.s~\eqref{eq:LLL} and \eqref{eq:OMEGAM0}, respectively), lower panel to the case in which
the fluctuations are included.
In both cases, a range of temperature where $\sigma$ decreases exists, that signals the partial restoration of
chiral symmetry (chiral symmetry cannot be restored exactly due to the soft explicit breaking in the action).

\begin{figure}[t!]
	\centering 
	\includegraphics[width=0.8\columnwidth]{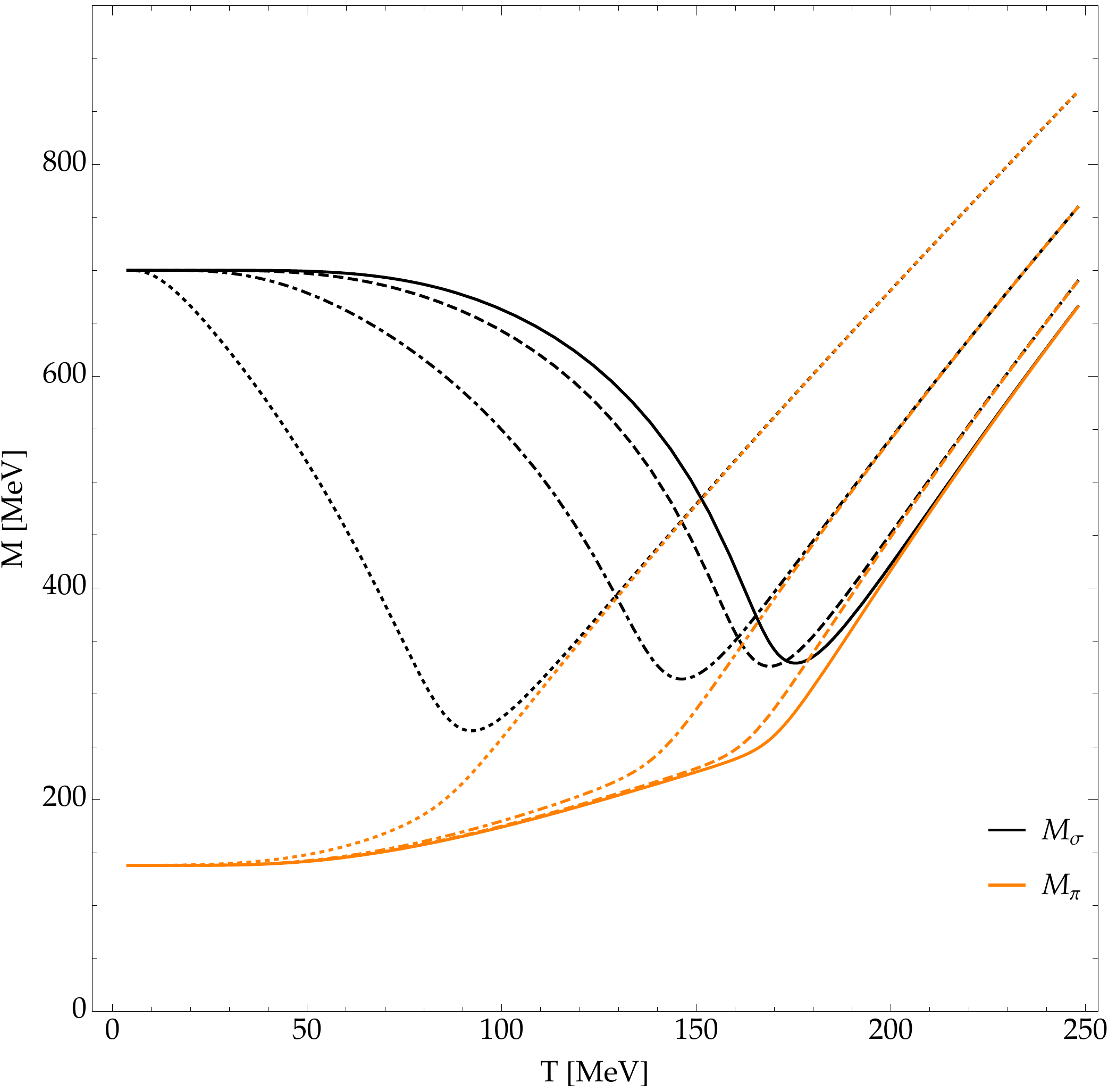}
	\caption{In-medium masses $M_\sigma$ (black) and $M_\pi$ (orange) as a function of $T$, 
	for several values of the chemical potential: $\mu=0$~MeV (continuous line), $\mu=100$~MeV (dashed), $\mu=200$~MeV (dot-dashed) and $\mu=300$~MeV (dotted).  Case with mesonic fluctuations.} 
	\label{fig:massaC2}
\end{figure}

In Fig.~\ref{fig:massaC2} we plot the in-medium masses of the $\sigma$-meson and pions as a function of temperature,
for several values of the quark chemical potential. These have been computed for the model with fluctuations included.
We notice that for each of the values of $\mu$ considered, a range of temperature exists in which the $\sigma$-meson mass
decreases while the pions mass increases, and the two match at high temperature signaling the approximate restoration
of the $O(4)$ symmetry, as well as the decoupling of these particles from the low energy spectrum of the model.
Moreover, the lowering of $M_\sigma$ to a minimum is a sign that the fluctuations of the scalar field are enhanced near the
chiral crossover.

\begin{figure}[t!]
	\centering 
	\includegraphics[width=0.8\columnwidth]{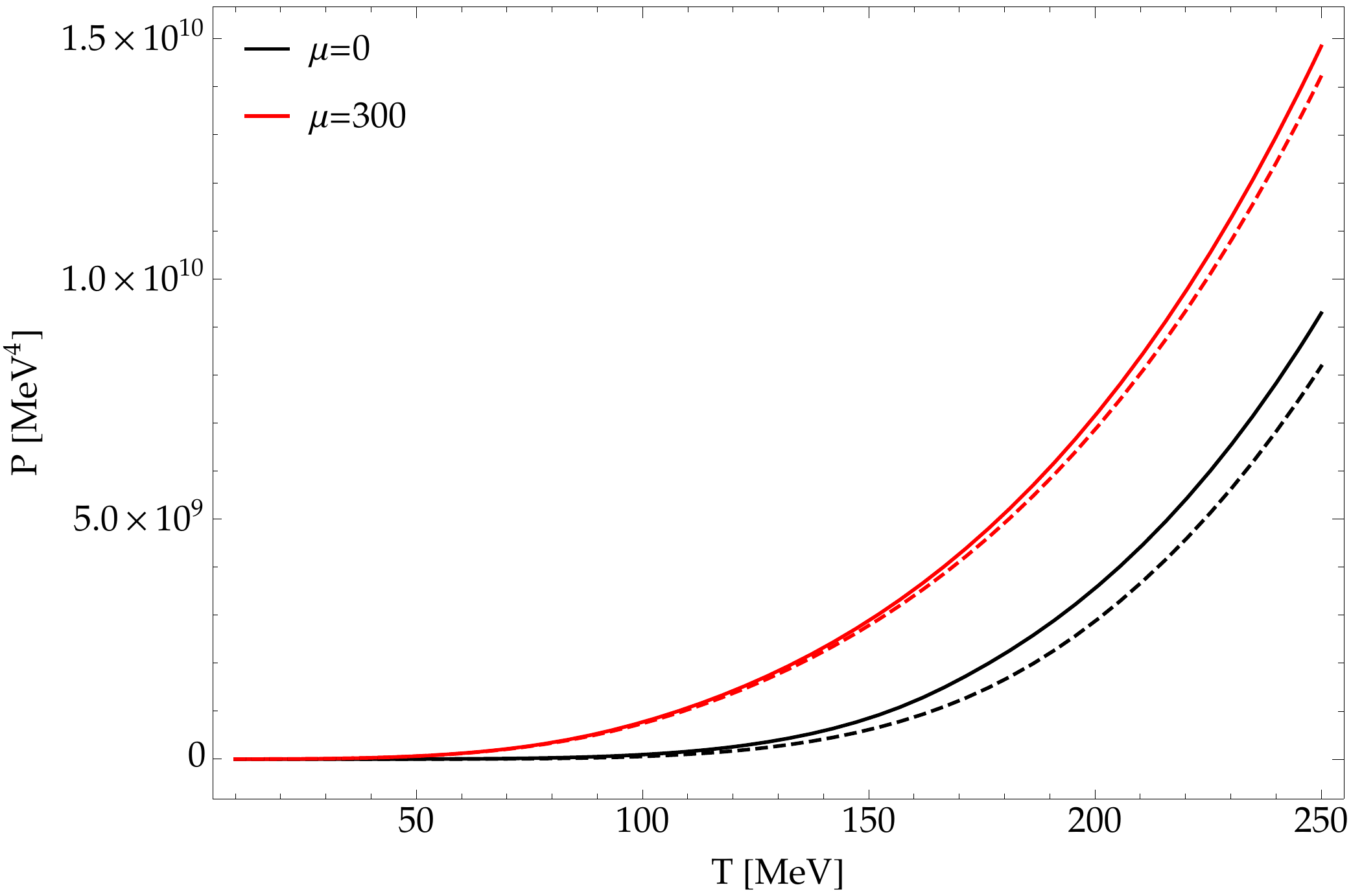}
	\caption{Pressure versus temperature for the models with (solid lines) and without (dashed lines) meson fluctuations,
	for $\mu=0$ (black lines) and $\mu=300$ MeV (red lines).}
	\label{fig:RTaBB}
\end{figure}

In Fig.~\ref{fig:RTaBB} we plot the pressure versus the temperature for the models with and without fluctuations,
for $\mu=0$ (black lines) and $\mu=300$ MeV (red lines). At fixed  $\mu$ and $T$  the fluctuations increase the pressure
as expected; however, we notice that at large values of $\mu$ the contribution of the fluctuations becomes
less important in comparison with the mean field pressure. 

\subsection{The thermodynamic curvature}

\begin{figure}[t!]
	\centering 
	\includegraphics[width=0.8\columnwidth]{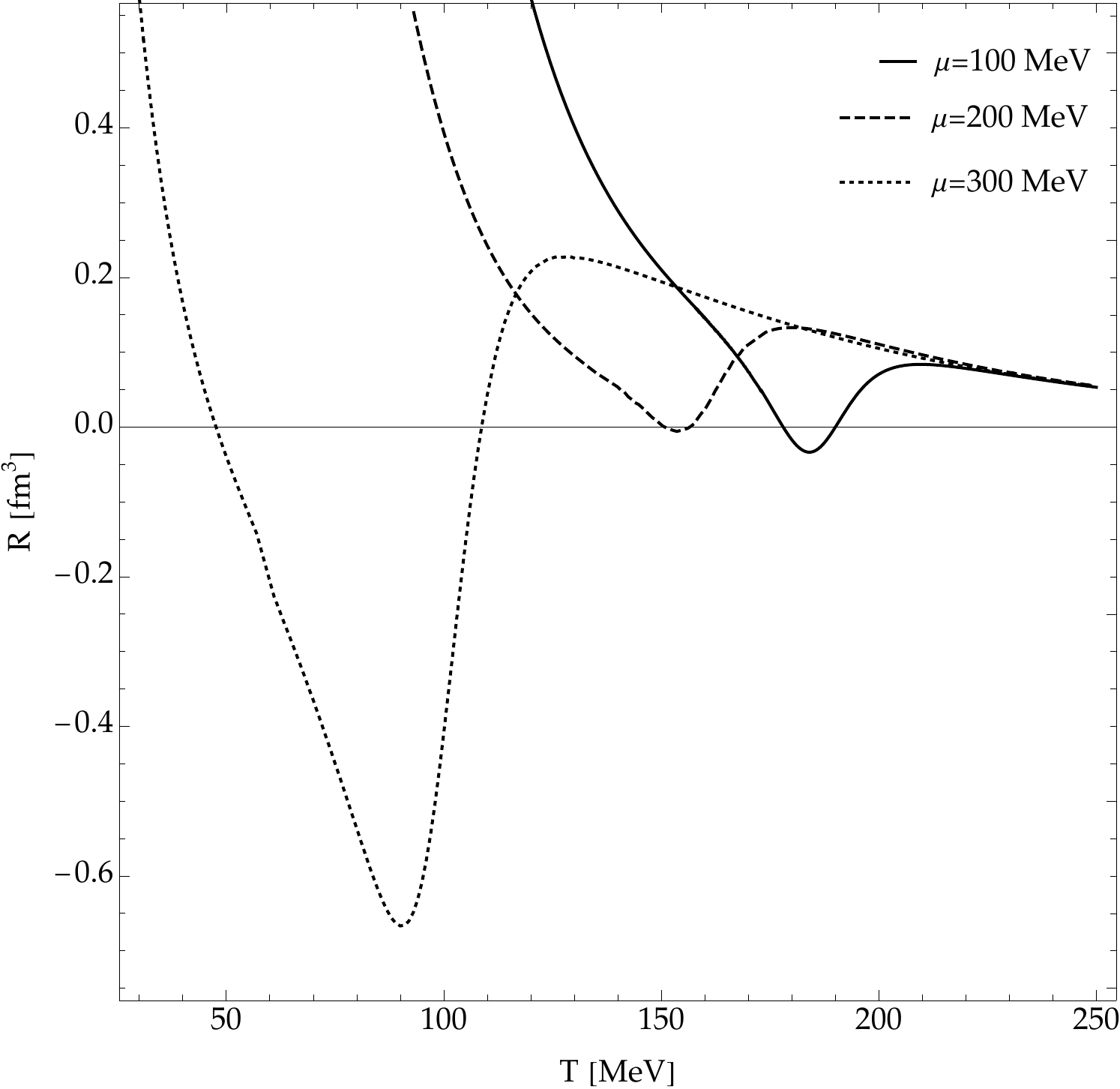}\\
	\includegraphics[width=0.8\columnwidth]{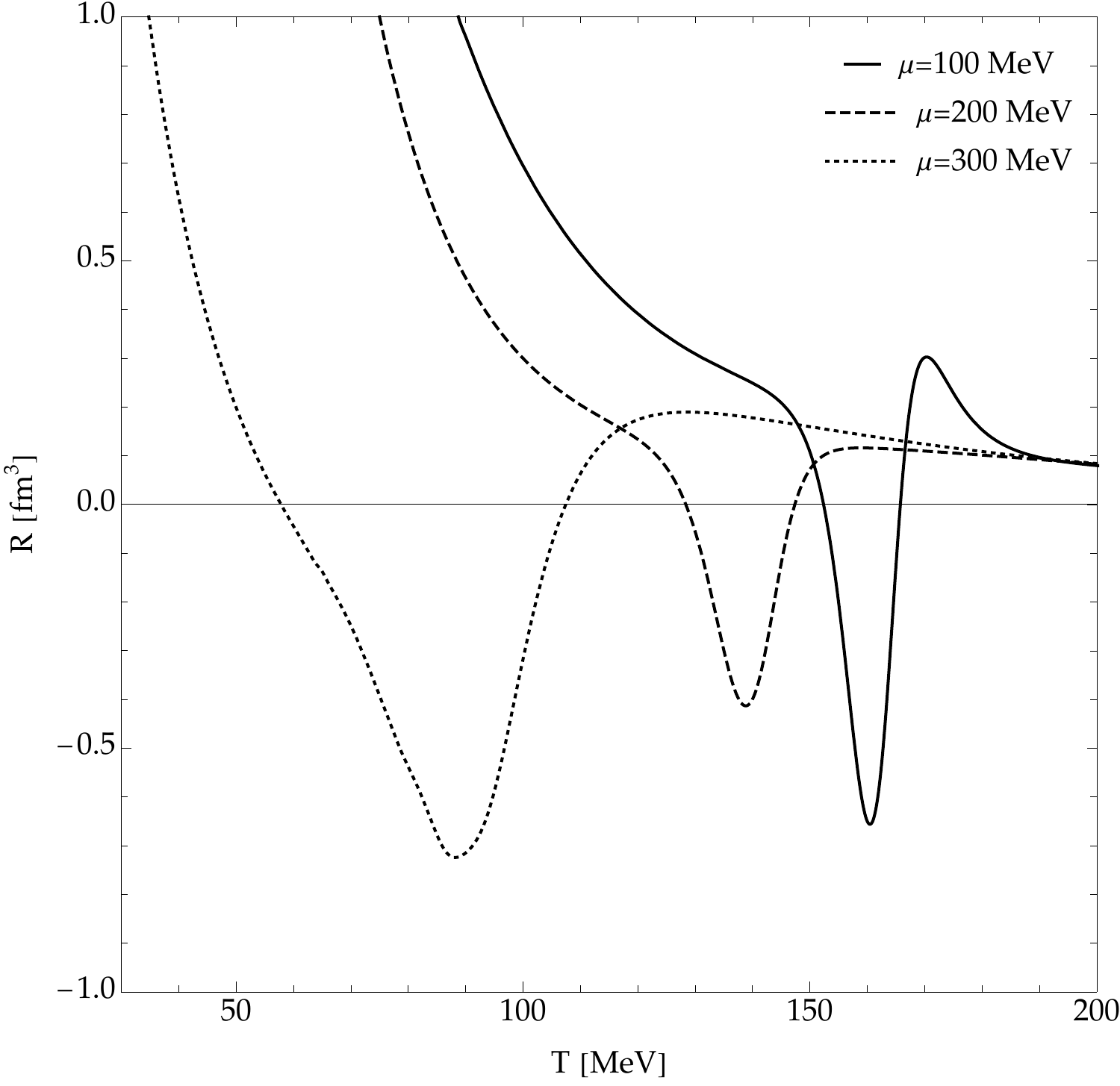}
	\caption{Scalar curvature, $R$, as a function of $T$ for $\mu=100$~MeV (continuous), $\mu=200$~MeV (dashed) and $\mu=300$~MeV (dotted).
	Upper and lower panels correspond to the cases without and with mesonic fluctuations.}
	\label{fig:RTa}
\end{figure}

In Fig.~\ref{fig:RTa} we plot the scalar curvature, $R$, versus temperature
for three values of the quark chemical potential: the upper panel corresponds to the case without fluctuations,
the lower panel to that with fluctuations.
We notice that in both cases, $R$ develops a groove around the chiral crossover,
in agreement with \cite{Castorina:2019jzw,Zhang:2019neb}. This is expected 
thanks to the relation between $R$ and the correlation volume around a phase transition:
as a matter of fact, at a second order phase transition
$R$ diverges due to the divergence of the correlation volume, while at a crossover the correlation length 
increases but remains finite
and susceptibilities are enhanced so $|R|$ is expected to grow up in the pseudocritical region.
Therefore, the thermodynamic curvature brings information about the correlation volume also near a crossover.

\begin{figure}[t!]
	\centering 
	\includegraphics[width=0.8\columnwidth]{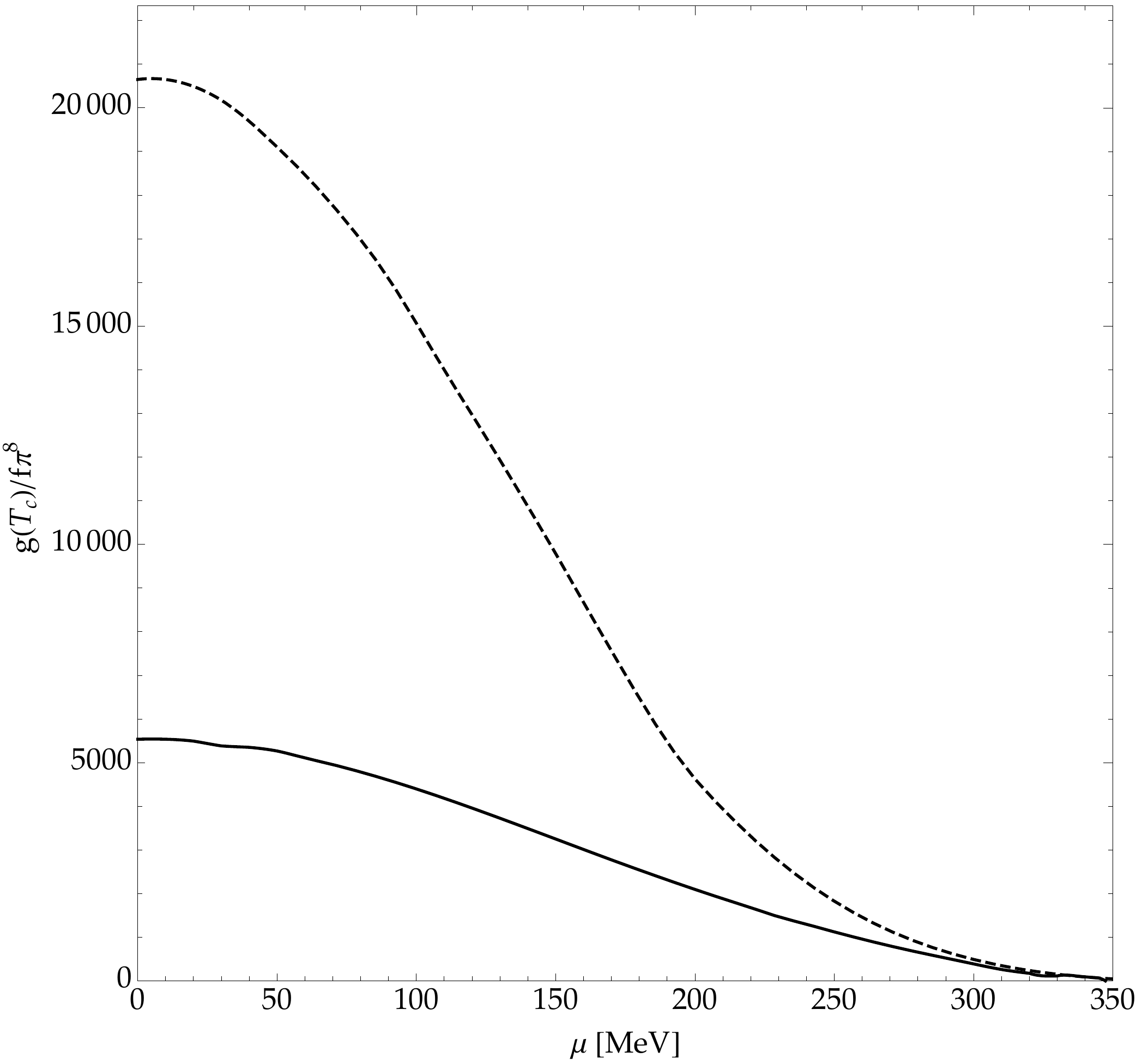}
	\caption{Determinant of the thermodynamic metric versus $\mu$ computed at the chiral crossover temperature 
	obtained as $T_{{\mathrm TG}}(\mu)$:
	solid line corresponds to the case with fluctuations while dashed line to the mean field thermodynamics.}
	\label{fig:RTaBBdd}
\end{figure}

In addition to this, we find that at small $\mu$ the values of $|R|$ are more pronounced when the fluctuations are included.
This is an interesting, new observation about the thermodynamic geometry
and is related to the fact that fluctuations make the chiral broken phase more unstable.
This can be seen from the determinant of the thermodynamic metric, $g$, see Fig.~\ref{fig:RTaBBdd}:
in the figure we define $T_\mathrm{TG}$ as the temperature at which $R$ develops its local minimum,
which is in agreement with other definitions, see also Fig.~\ref{fig:TcC2}.
At small $\mu$ in the critical region the determinant with fluctuations is smaller than the one without fluctuations
($g=0$ corresponds to thermodynamic instability and infinite curvature),
while increasing $\mu$ the determinant in the critical region is not very affected by the presence of the fluctuations.
This is in line with the results of the pressure in Fig.~\ref{fig:RTaBB} in which we show that
fluctuations do not give a substantial contribution in the critical region at large $\mu$. 
When $\mu$ is large enough, 
$R$ is enhanced in the critical region both with and without fluctuations (see also Fig.~\ref{fig:TcC2aaa} below).
This is most likely related to the fact that the critical endpoint with the second order phase transition
and the divergent correlation length already appears within the mean field approximation, 
so the main role of the fluctuations is that to change the critical exponents but not to change the phase structure.

The scalar curvature changes sign around the crossover, both with and without fluctuations:
this is in agreement with \cite{Castorina:2019jzw,Zhang:2019neb} and can be interpreted as
a rearrangement of the collective interactions in the hot medium around the chiral crossover,
from statistically repulsive (due to the fermionic nature of the bulk) to attractive.
This piece of information was not accessible to previous model calculations on the QCD phase diagram
and represents a merit of the thermodynamic geometry.

\subsection{The critical temperature and the endpoint}

\begin{figure}[t!]
	\centering 
	\includegraphics[width=0.8\columnwidth]{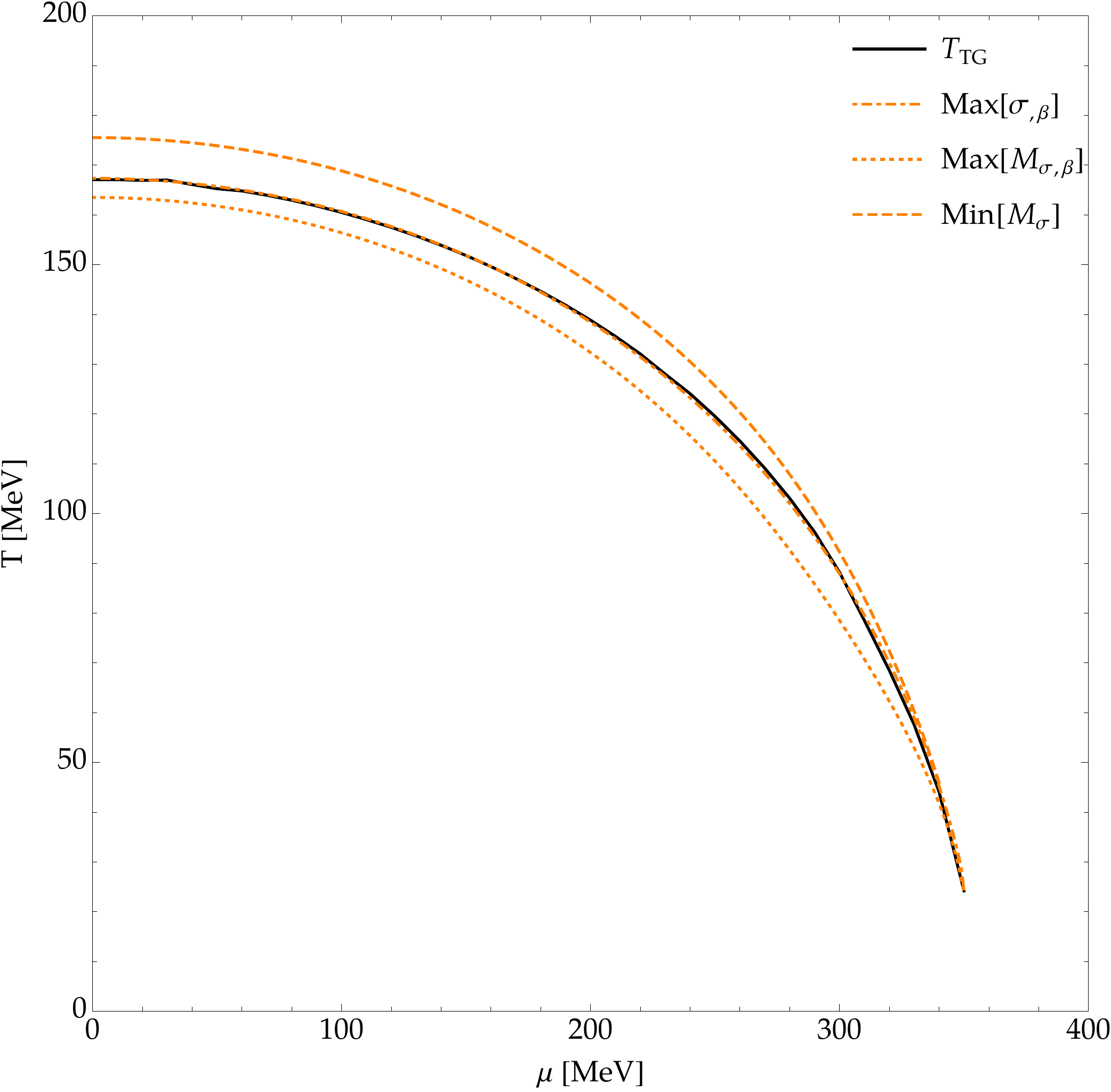}
	\caption{Crossover temperature versus $\mu$ obtained with four definitions:
	from the maximum of $M_{\sigma,\beta}$ (orange dotted line), from the maximum of 
	$\sigma_{,\beta}$ (orange dot-dashed line), from the minimum of $M_\sigma$ (orange dashed line) 
	and from the local minimum  of $R$ (black line). Case with mesonic fluctuations.}
	\label{fig:TcC2}
\end{figure}

The crossover nature of the transition to the chiral symmetric phase at high temperature
leaves an ambiguity on the definition of a critical temperature: in fact, it is possible to adopt several definitions
to identify the critical region, in which the order parameter decreases substantially. 
We compare the predictions of the model using four different definitions. 
Firstly, we define the pseudocritical temperature, $T_c(\mu)$, as the temperature corresponding to the maximum of $\partial\sigma/\partial\beta$ (which coincides with the maximum of $\partial\sigma/\partial\gamma$).
A second definition is the temperature at which $\partial M_\sigma/\partial\beta$ is maximum (the same of $\partial M_\sigma/\partial\gamma$). 
Thirdly, we can define $T_c$ as the one at which $M_\sigma$ is minimum (since at this temperature
the correlation length of the fluctuations of the order parameter is the largest). 
Finally, the peculiar structure of $R=R(T)$ at a given $\mu$ 
allows for the fourth definition, namely the temperature at which
$R$ presents its local minimum.: we denote this by $T_\mathrm{TG}$.

In Fig.~\ref{fig:TcC2} we show $T_c$ versus $\mu$ obtained with the four definitions.
We notice that the different definitions give consistent results with each other. 
This supports the idea that we can use the local minima of
$R$ to identify the chiral crossover, which in turn suggests that $R$ is sensitive to the crossover from the broken to the unbroken phase
even though this is not a real second order phase transition.  

\begin{figure}[t!]
	\centering 
	\includegraphics[width=0.8\columnwidth]{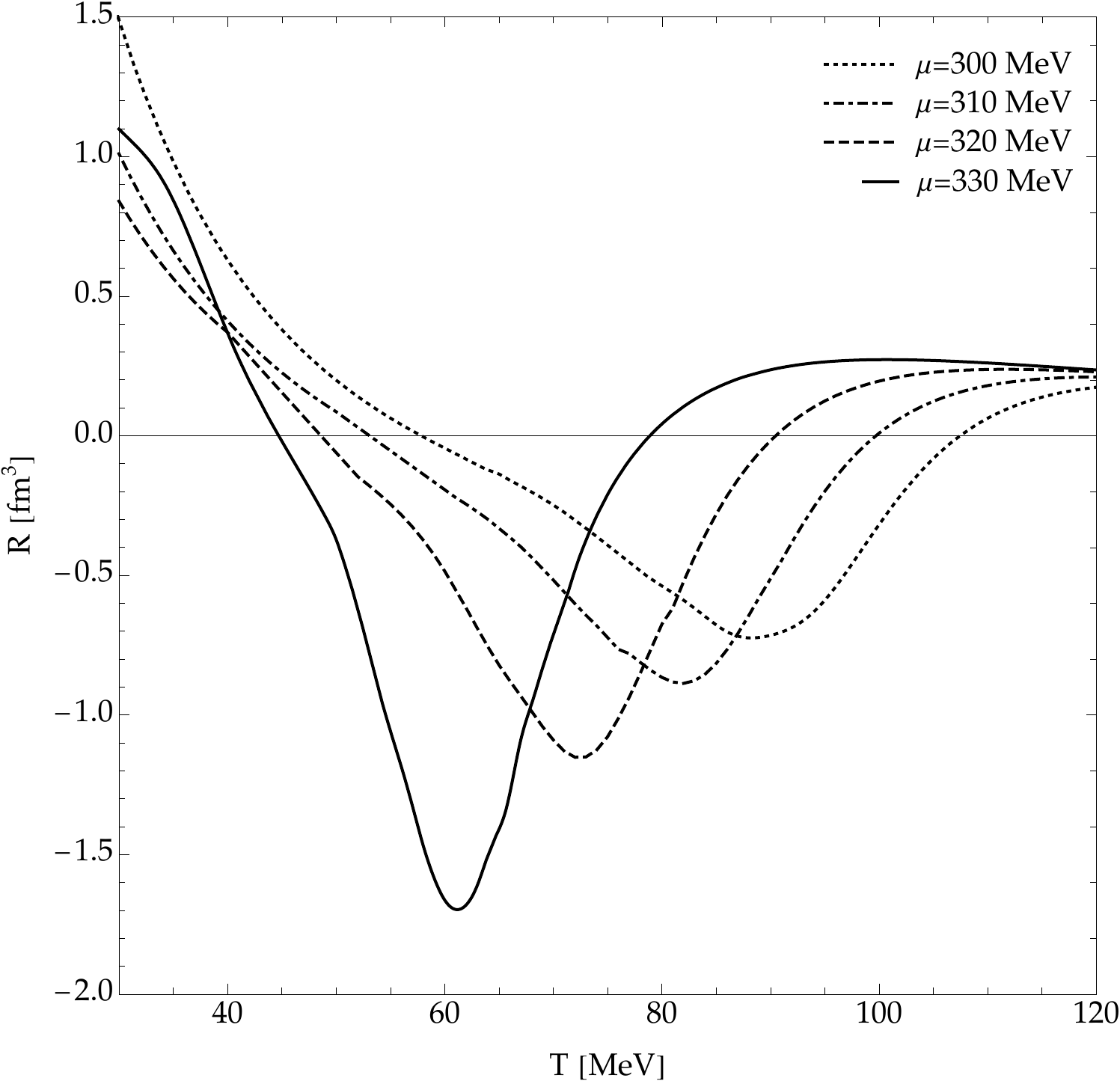}
	\caption{Thermodynamic curvature versus temperature for several values of $\mu$ close to the critical endpoint. 
	 Case with mesonic fluctuations.}
	\label{fig:TcC2aaa}
\end{figure}

In the phase diagram shown in Fig.~\ref{fig:TcC2} the crossover line terminates at a critical endpoint, $\mathrm{CEP}$,
located at $(\mu_\mathrm{CEP},T_\mathrm{CEP})=(350~\mathrm{MeV},30~\mathrm{MeV})$.  
Approaching this point along the critical line, the crossover turns into a second order phase transition with divergent
susceptibilities, then
the transition becomes first order with jumps of the condensate across the transition line.

In Fig.~\ref{fig:TcC2aaa} we plot $R$ versus temperature for values of $\mu$ close to the 
critical endpoint (the case without fluctuations is not shown because it is very similar).
As expected, approaching the critical endpoint the magnitude of $|R|$ becomes larger, as it should be since the crossover
becomes a second order phase transition there and $R$ should diverge at the CEP.

\subsection{Thermodynamic curvature and correlation volume}  
  
\begin{figure}[t!]
	\centering 
	\includegraphics[width=0.8\columnwidth]{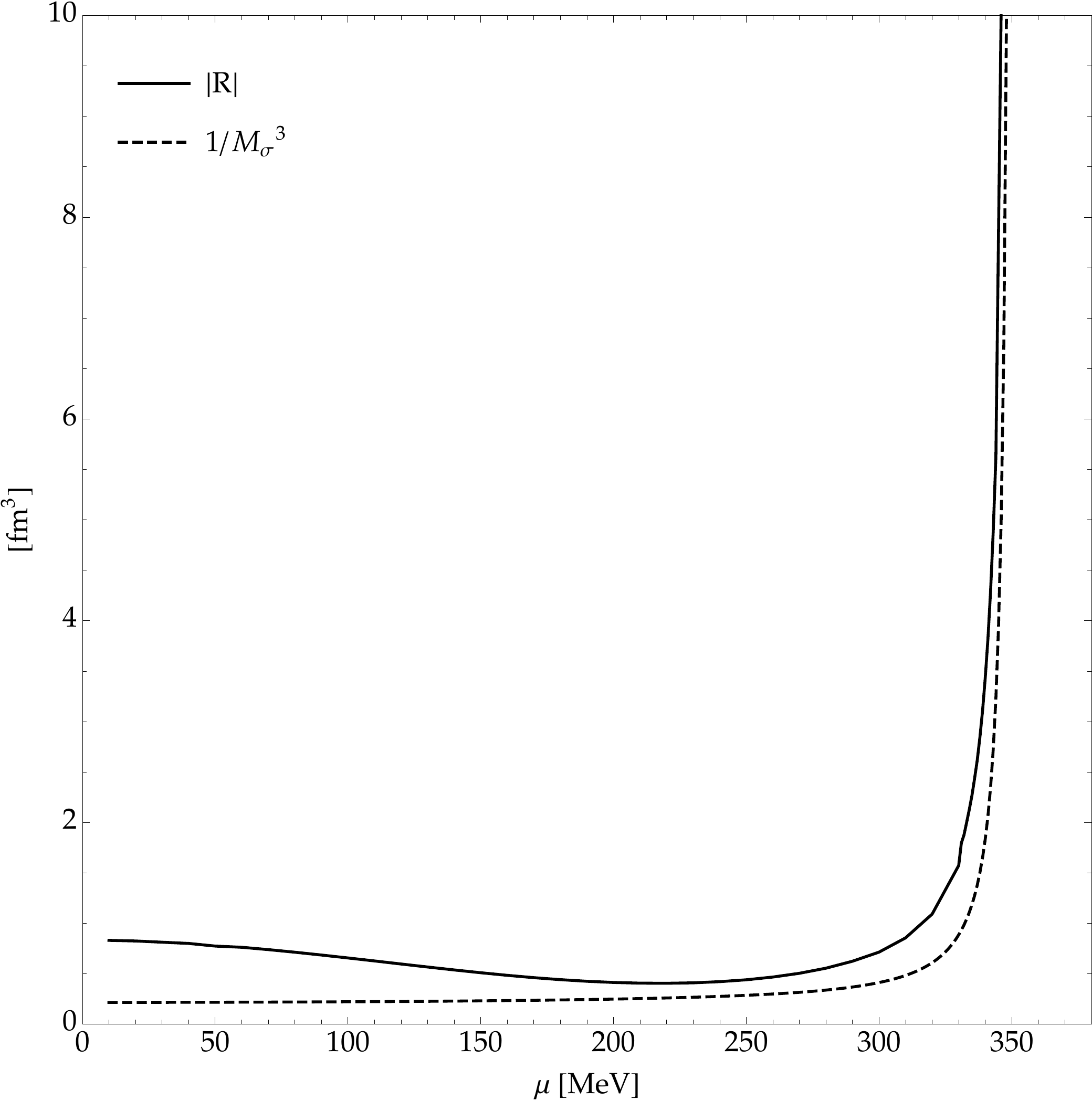}
	\caption{Thermodynamic curvature versus $\mu$ at the critical line, 
	compared with the inverse of the correlation volume $1/M_\sigma^3$. Case with mesonic fluctuations.}
	\label{fig:TcC2aaaBBB}
\end{figure}

It is interesting to compare the thermodynamic curvature around the critical line,
with the correlation volume $\xi^3$, where $\xi$ is the correlation length.
This comparison is interesting since according to hyperscaling arguments,
around a second order phase transition  $|R| = K\xi^3$ with $K$ of the order of unity;
restoration of chiral symmetry is a crossover rather than a real phase transition, at least far from the critical endpoint,
therefore we can check how the hyperscaling relation works around such a smooth crossover and how it changes
approaching the CEP.

In Fig.~\ref{fig:TcC2aaaBBB} we compare the thermodynamic curvature, computed along the critical line,
with the correlation volume, the latter being estimated by taking $\xi=1/M_\sigma$ as a measure of the correlation length
of the fluctuations of the order parameter. 
We find that both the correlation volume and the thermodynamic curvature behave qualitatively
in the same way near the CEP; moreover, the numerical values of the two quantities is comparable in the critical region.
We conclude that our study supports the idea that $|R|=K\xi^3$ in proximity of the second order phase transition.

In the small $\mu$ regime the relation between the curvature and the correlation volume
does not need to be satisfied since in this regime the critical line is a smooth crossover.
In fact, we find that for small values of $\mu$ the agreement between $|R|$ and $\xi^3$ 
is not as striking as the one in proximity of the CEP; nevertheless, we still find that the two quantities
behave qualitatively in the same way, namely they stay approximately constant for a broad range of $\mu$
then grow up as the CEP is reached.


\section{\label{sec:SC}Conclusions}
We have studied the thermodynamic geometry 
around the chiral phase transition at finite temperature and chemical potential.
The phase transition has been studied within the quark-meson (QM) model augmented with meson fluctuations;
within this model the phase transition at large temperature and small chemical potential is actually
a smooth crossover, which turns to a second order phase transition at the critical endpoint
then becomes a first order phase transition at large values of the chemical potential.

The main goals have been to analyze the relation between the thermodynamic curvature, $R$,
and the correlation volume for smooth crossovers, and how this changes approaching a second order phase transition
at the critical endpoint. Moreover, we have studied the effect of the fluctuations, pions and $\sigma-$meson, 
on the top of the mean field
thermodynamics and how these affect the thermodynamic curvature around the crossover.
Of particular interest is the $\sigma-$meson since it corresponds to the amplitude fluctuation mode
and its mass can be related directly to the correlation length of the fluctuations of the order parameter. 
Fluctuations have been introduced within the Cornwall-Jackiw-Toumbulis effective potential formalism \cite{Cornwall:1974vz}
in the Hartree approximation;
we have neglected the zero point energy contributions of the meson fields, both for the sake of simplicity and to avoid
the unexpected behavior of thermodynamic quantities when these are included and two renormalization scales are 
needed \cite{Zacchi:2017ahv}.
Within this approach, the Schwinger-Dyson equations for the meson propagators become simple equations
for the meson masses that can be solved, consistently with the gap equations, to get the condensate
and the masses as a function of temperature and chemical potential.
This study is a natural continuation of previous works \cite{Castorina:2019jzw,Zhang:2019neb}
in which the same problem has been analyzed within the mean field approximation.

We have found that in the region of small values of $\mu$, the fluctuations enhance the magnitude of the curvature.
We understand this in terms of the stability of the phase with broken chiral symmetry, 
that can be analyzed by  the determinant of the metric, $g$:
in fact, the condition of stability reads $g>0$ while $g=0$ corresponds to a phase boundary where
a phase transition happens and $R$ diverges, so the smaller the $g$ the closer the system is
at a phase transition and the larger is $R$. 
We have found that the determinant with fluctuations and around the crossover is smaller 
than $g$ without fluctuations in the same range of $T$ and $\mu$, meaning that fluctuations
make the chiral broken phase less stable. This result is expected, since fluctuations of the order parameter
represented by the $\sigma-$meson tend to wash out the $\sigma-$condensate.

On the oher hand, at larger values of $\mu$ and in proximity of the critical endpoint, 
the fluctuations do not bring significant changes to the mean field solution
around the critical line and $R$ is less sensitive to the fluctuations. 
This is also easy to understand, because the mean field thermodynamics already predicts
the existence of the critical endpoint with a divergent curvature \cite{Castorina:2019jzw,Zhang:2019neb},
so the role of the fluctuations is just that to change the mean field critical exponents.

We have found that $R$ changes sign in the pseudocritical region: this suggests that around the chiral crossover,
the interaction changes at mesoscopic level from being statistically repulsive to attractive.
This change in the nature of the interaction is not accessible to methods based on standard thermodynamics, 
and this prediction represents one of the merits of the thermodynamic geometry.

We have verified that in the critical region around the critical endpoint $|R|$ scales with the correlation volume,
$|R|=K\xi^3$ with $K=O(1)$, in agreement with hyperscaling arguments: thus $|R|$ brings information
on the correlation volume. In proximity of the crossover at small $\mu$ the correspondence between $|R|$
and the correlation volume is not as good as the one we have found at large $\mu$, which is not surprising
because at small $\mu$ the chiral crossover is quite smooth; nevertheless, we have found that $R$ develops a characteristic
groove structure with a pronounced local minimum, 
suggesting that it is capable to capture  the pseudocritical behavior of the condensate.

This study presents the natural continuation of the work started in \cite{Castorina:2019jzw,Zhang:2019neb},
and offers possibilities for further investigations. From the fields theory point of view, two natural questions are 
whether it is worth to relax the Hartree approximation to get a more realistic description of the quantum fluctuations, 
and whether the renormalization of the zero point energy of mesons has to be done on the same
footing of that of quarks. Investigating these topics might bring to a more confident application of the theory
of quantum fluctuations to the thermodynamic geometry. Moreover, fluctuations can be treated also with the
Functional Renormalization Group (FRG) approach \cite{Wetterich,Skokov:2010wb} under suitable assumptions on the
full effective potential: it is of a certain interest to study the effects of the fluctuations on the thermodynamic
geometry using FRG. We plan to report on  these topics in the near future.

\begin{acknowledgments}	
	M. R. acknowledges John Petrucci for inspiration. 
	M. R. is supported by the National Science Foundation of China (Grants No.11805087 and No. 11875153)
	and by the Fundamental Research Funds for the Central Universities (grant number 862946).
\end{acknowledgments}

\end{document}